\documentclass[11pt]{article}  
\usepackage{menuproc}
\usepackage{cite}
\usepackage{epsfig}
\usepackage{amsmath,amssymb}
\def\bbox#1{\mbox{\boldmath {$#1$}}}
\begin{document}
%
%
%
\titlematter{Covariant meson-exchange model of the $\bar K N$
interaction}%
{A.D. Lahiff}%
{TRIUMF, 4004 Wesbrook Mall, Vancouver, B.C., Canada V6T 2A3}
{A covariant meson-exchange model of the $\bar K N$ interaction
within the framework of the Bethe-Salpeter
equation is presented. With just one free parameter we are able
to get a good description of the available experimental data from
below threshold to 300 MeV laboratory momentum.}

%
%


The construction of nonperturbative models of the $\bar K N$ interaction
has a long history, dating back to early work such as that of
Dalitz, Wong, and Rajasekaran~\cite{dalitz:1967}. This model consisted of
a coupled-channel Schr\"{o}dinger equation with static vector meson
exchange potentials, and was able to dynamically generate
the S-wave $\Lambda (1405)$ resonance. More recent calculations
are based on similar ideas, and commonly use a coupled-channel 
Lippmann-Schwinger equation to iterate a set of potentials to infinite 
order, with the strengths of the potentials in the various
channels constrained by SU(3) symmetry. The strong attraction produced
by the $I=0$ $\bar K N \rightarrow \bar K N$ potential results in the
dynamical generation of the $\Lambda (1405)$ resonance as an unstable
$\bar K N$ bound state. Further resonances, including the $\Lambda (1670)$
and $\Sigma (1620)$, have also been found to be formed 
dynamically~\cite{Siegel:1999ac,Oset:2001cn}.

Previous works have in general relied upon non-relativistic formulations,
or have made use of the (ladder) Bethe-Salpeter equation~\cite{Salpeter:1951sz} 
but solved it in an approximate way. Here we outline a covariant model of
low-energy $\bar K N$ scattering based on the 4-dimensional Bethe-Salpeter
equation, and present some preliminary results.

The multi-channel Bethe-Salpeter equation for the $\bar K N$ system is
\begin{equation}
T_{nm}(q',q;P)  =  V_{nm}(q',q;P)  
- \sum _k \frac{i }{ (2 \pi )^4 }\int d^4 q'' 
V_{nk}(q',q'';P)G_k(q'';P) 
 T_{km}(q'',q;P)  , \label{eq:bse}
\end{equation}
where $m$ ($n$) label the initial (final) states, and
$k$ is summed over the included channels 
($K ^- p$, $\bar K ^0 n$, $\Lambda \pi ^0$,
$\Sigma ^- \pi ^+$, $\Sigma ^0 \pi ^0$, $\Sigma ^+ \pi ^-$,
$\Lambda\eta$, and $\Sigma ^0 \eta$). Also,
$P=(\sqrt{s},\bbox{0})$ is the total 4-momentum in the center-of-mass (c.m.), 
while $q$, $q'$ and $q''$ are the relative 4-momenta in the
initial, final and intermediate states.
The two-body propagator $G_k(q;P)$ is given by the product of the
appropriate baryon and pseudoscalar meson propagators.
The interaction kernels $V_{nm}$ are constructed from the $s$- and
$u$-channel baryon poles and $t$-channel vector meson
pole diagrams obtained from the usual SU(3)-symmetric $BBP$, $BBV$,
and $PPV$ interaction Lagrangians~\cite{deSwart:1963gc} (here $B$, $P$, and $V$ represent the
$J^P = 1/2 ^+$ baryons, the pseudoscalar mesons, and the vector mesons,
respectively). In order to regularize the
Bethe-Salpeter equation all the propagators are multiplied by form factors, which are given by
\begin{equation}
f_{B_k} (p^2)  = \left( \frac{m_{B_k}^2 - \Lambda ^2}{p^2 - \Lambda ^2 + i \epsilon } \right) ^2  , \hspace*{0.4cm}
f_{P_k} (p^2)  = \left( \frac{m_{P_k}^2 - \Lambda ^2}{p^2 - \Lambda ^2 + i \epsilon } \right) ^2  , \hspace*{0.4cm}
f_{V_k} (t)  = \left(\frac{ - \Lambda ^2}{t - \Lambda ^2 + i \epsilon } \right) ^2   . \hspace*{0.4cm}
\label{eq:ffdef}
\end{equation}
We use the same cutoff mass $\Lambda$ for all particles in order to 
minimize the number of free parameters.

The method of solution is the same as that described in 
Ref.~\cite{Lahiff:1999ur}.
A partial wave decomposition is applied to Eq.~(\ref{eq:bse}) which gives
a system of coupled 2-dimensional integral equations. The singularities
in the relative-energy variables are handled by performing a
Wick rotation~\cite{Wick:1954eu}, i.e., the relative-energy integration
contour is rotated from the real to the imaginary axis. 
All of the basic coupling constants in the model are fixed using information from 
other sources, such as decay widths and vector meson dominance (VMD),
and are not left as
free parameters. The only adjustable parameter is the cutoff mass,
which we fix by fitting to the $\bar K N$ data. The values of the basic
coupling constants and the cutoff mass are shown in Table 1.

\begin{table}[h]
\parbox{.55\textwidth}{\begin{tabular}{lll} \hline \hline 
coupling constants & & \\ \hline
$g_{\pi\pi\rho}$ & 6.05 & $\Gamma (\rho ^0 \rightarrow \pi ^+ \pi ^- )$ \\
$g_{N N \rho}$ & 2.52 & $\Gamma ( \rho ^0 \rightarrow e^+ e^- ) $ \\ 
$\kappa _{N N \rho}$ & 3.71 & VMD \\
$g_{N N \omega }$ & 3.4 $g_{N N \rho}$ & 
$\Gamma ( \rho ^0 \rightarrow e^+ e^-) / 
\Gamma ( \omega \rightarrow e^+ e^-) $ \\
$\kappa _{N N \omega}$ & -0.12 & VMD \\
$f_{N N \pi}^2/ 4 \pi$ & 0.075 &  nucleon-nucleon data \\ \hline 
$F/(F+D)$ ratios & & \\ \hline 
$\alpha _{PV}$ & 0.4 &  semileptonic hyperon decays \\
$\alpha _V^e$ & 1.0 & universality \\
$\alpha _V^m$ & 0.28 & relativistic SU(6) \\ \hline 
cutoff mass & & \\ \hline
$\Lambda$ & 2.42 GeV & \\ \hline \hline 
\end{tabular}} \hfill
\parbox{.30\textwidth}{\caption{Parameters of the model. The basic
coupling constants and the $F/(F+D)$ ratios are fixed: the sources
of the values used are given in the right-hand column. Also note that
we assume ideal $\phi - \omega$ mixing, and take the physical $\eta$
to be the pure octet state.}}
\end{table}

The $K ^- p$ cross sections are shown in Figure 1, 
where we find good
agreement with the experimental data. The $Y \eta$ channels give
non-negligible contributions, even though the energies
we consider are well below the $\Lambda \eta$ and $\Sigma \eta$
thresholds.

\begin{figure}[ht]
\parbox{.50\textwidth}{\epsfig{file=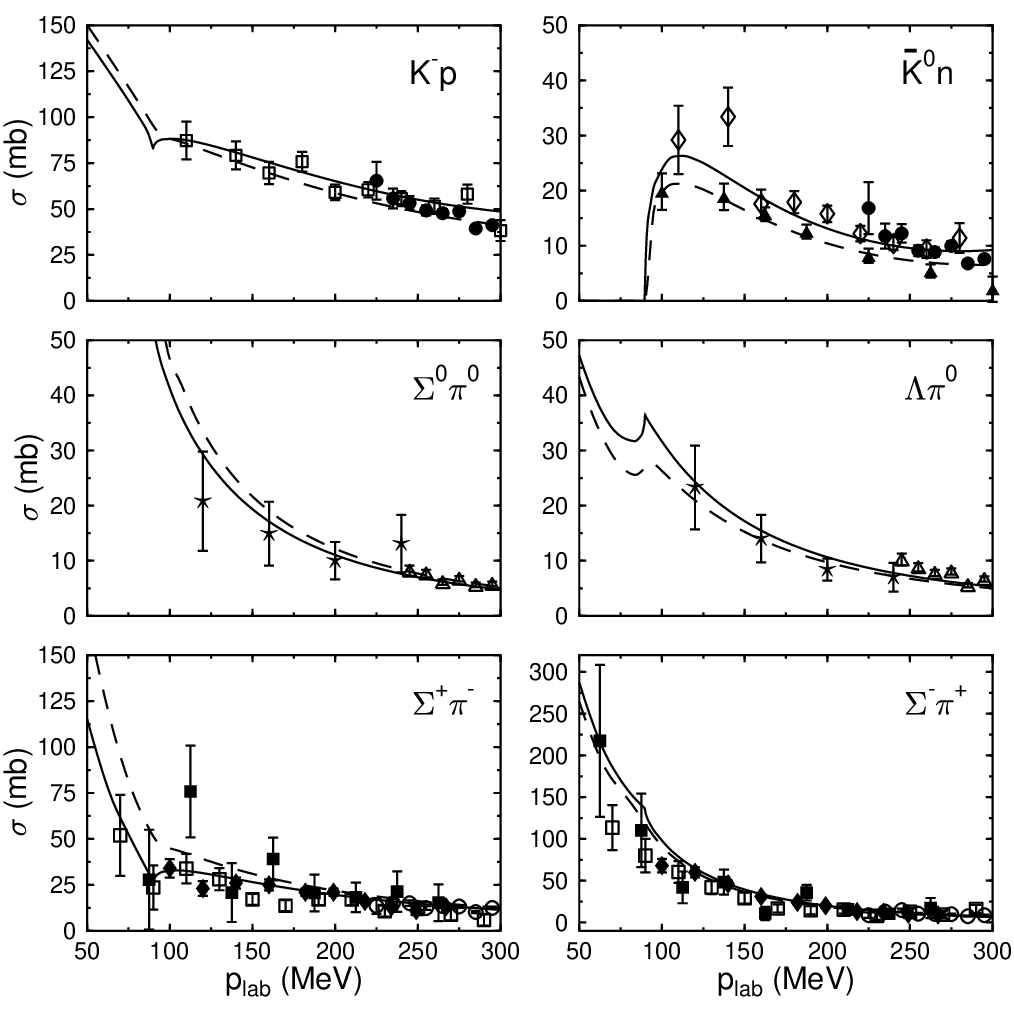,width=8.0cm}}
\hfill
\parbox{.50\textwidth}{\epsfig{file=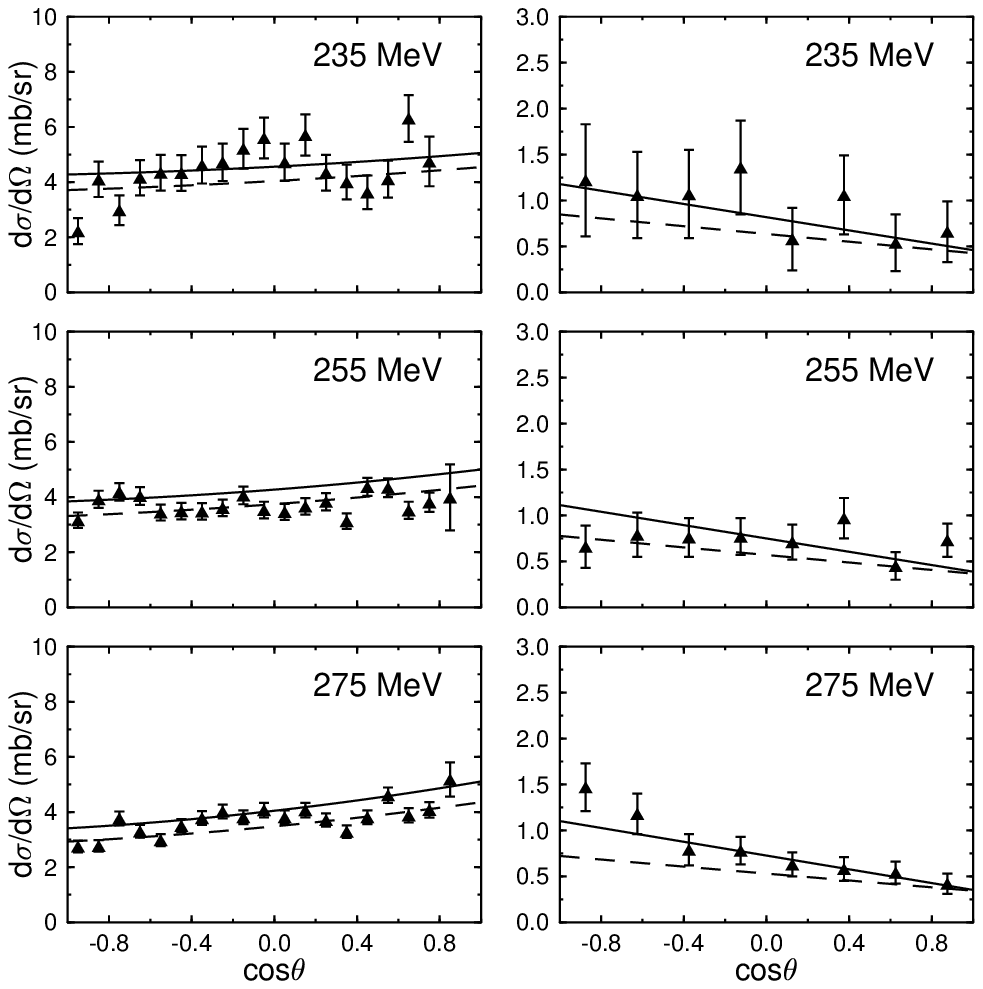,width=8.0cm}}
\caption{The first two columns show the cross sections for
the six final states. The third and fourth columns show the
elastic and charge-exchange differential cross sections,
respectively, compared to the experimental data of
Mast et al.~\protect\cite{Mast:1976pv}.
The solid lines correspond to the full model, while the
$Y \eta$ channels were omitted in the calculations giving the
dashed lines.}
\label{fig:kcps}
\end{figure}

We now turn to the threshold behavior.
For the $K^- p$ scattering length we find 
$a_{K^- p} = -0.54 + i1.2$ fm, which is consistent with the
experimental value of $a_{K^- p} = -0.78 \pm 0.18 + i(0.49 \pm 0.37)$ fm
obtained in a kaonic hydrogen experiment~\cite{Iwasaki:1997wf}.
The relative strengths of
the different channels at threshold are tightly constrained by the 
threshold branching ratios $\gamma$, $R_c$, and $R_n$, which are 
given in Refs.~\cite{Nowak:1978au,Tovee:1971ga} as
\[
\gamma = \frac{ \Gamma (K^-p \rightarrow \Sigma ^- \pi ^+ )}{
           \Gamma (K^-p \rightarrow \Sigma ^+ \pi ^- )} = 2.36 \pm 0.04  , \hspace*{0.5cm}
R_c = \frac{ \Gamma (K^-p \rightarrow \mbox{charged} ) }{
           \Gamma (K^-p \rightarrow \mbox{all} )} = 0.664 \pm 0.011  , \hspace*{0.5cm} 
\]   	   
\[	    	   
R_n = \frac{ \Gamma (K^-p \rightarrow \Lambda \pi ^0 ) }{
           \Gamma (K^-p \rightarrow \mbox{neutral} )} = 0.189 \pm 0.015  . 
\]
Our values for the branching ratios are $\gamma = 2.14$,  $R_c = 0.651$, 
and $R_n = 0.132$, 
which are in reasonable agreement with the experimental values. 
As found previously by
Oset and Ramos~\cite{Oset:1998it}, the $Y \eta$ channels 
(in particular $\Lambda\eta$)
give important contributions to the branching ratios. When
the $Y \eta$ channels are neglected, $R_c$ and $R_n$ are essentially
unchanged, but $\gamma$ reduces to $1.38$.

Finally we consider the energy region below the $\bar K N$ threshold,
where the $I=0$ $\Sigma\pi\rightarrow\Sigma\pi$ amplitude exhibits the
$\Lambda (1405)$ resonance.
In Figure 2 we compare the $\Sigma\pi$ mass spectrum of the 
$\Lambda  (1405)$ obtained
in our model with experiment, and find good agreement.

\begin{figure}[t]
\parbox{.50\textwidth}{\epsfig{file=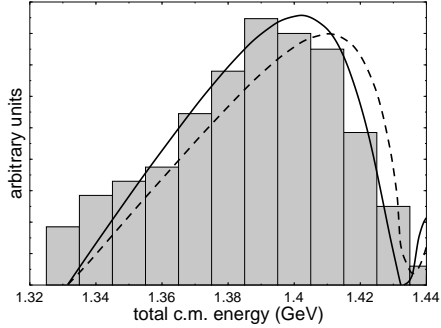,width=6.0cm}}
\hfill
\parbox{.42\textwidth}{\caption{The $\Sigma\pi$ mass distribution 
compared to the experimental data from 
Hemingway~\protect\cite{Hemingway:1985pz}. The solid line is the
result of the full calculation, while the dashed line shows the
effect of omitting the $Y \eta$ channels.}}
\label{fig:ms1405}
\end{figure}

To summarize, we have solved the multi-channel Bethe-Salpeter equation
for the $\bar K N$ system by means of a Wick rotation, and obtained good
agreement with the low-energy $\bar K N$ experimental data by adjusting
a single cutoff mass. Future work will include extending the model to
higher energies, and searching for evidence of additional 
dynamically-generated S- and P-wave hyperon resonances.

\acknowledgment{This work was supported in part by a grant from the 
Natural Sciences and Engineering Research Council of Canada.}



\begin{references}[99]   

\bibitem{dalitz:1967}
R.~H.~Dalitz, T.~C.~Wong, and G.~Rajasekaran,
Phys.\ Rev.\ {\bf 153}, 1617 (1967).

\bibitem{Siegel:1999ac}
P.~B.~Siegel and S.~Guertin,
PiN Newsletter\  {\bf 15}, 246 (1999).

\bibitem{Oset:2001cn}
E.~Oset, A.~Ramos and C.~Bennhold,
arXiv:nucl-th/0109006.

\bibitem{Salpeter:1951sz}
E.~E.~Salpeter and H.~A.~Bethe,
Phys.\ Rev.\  {\bf 84}, 1232 (1951).

\bibitem{deSwart:1963gc}
J.~J.~de Swart,
Rev.\ Mod.\ Phys.\  {\bf 35}, 916 (1963).


\bibitem{Lahiff:1999ur}
A.~D.~Lahiff and I.~R.~Afnan,
Phys.\ Rev.\ C {\bf 60}, 024608 (1999).

\bibitem{Wick:1954eu}
G.~C.~Wick,
Phys.\ Rev.\  {\bf 96}, 1124 (1954).

\bibitem{Mast:1976pv}
T.~S.~Mast {\it et al.},
Phys.\ Rev.\ D {\bf 14}, 13 (1976).


\bibitem{Iwasaki:1997wf}
M.~Iwasaki {\it et al.},
Phys.\ Rev.\ Lett.\  {\bf 78}, 3067 (1997).


\bibitem{Nowak:1978au}
R.~J.~Nowak {\it et al.},
Nucl.\ Phys.\ B {\bf 139}, 61 (1978).

\bibitem{Tovee:1971ga}
D.~N.~Tovee {\it et al.},
Nucl.\ Phys.\ B {\bf 33}, 493 (1971).



\bibitem{Oset:1998it}
E.~Oset and A.~Ramos,
Nucl.\ Phys.\ A {\bf 635}, 99 (1998).

\bibitem{Hemingway:1985pz}
R.~J.~Hemingway,
Nucl.\ Phys.\ B {\bf 253}, 742 (1985).

\end{references}
\end{document}